\documentclass[twocolumn,aps,prl,superscriptaddress,longbibliography]{revtex4-2}

\pdfoutput=1

\usepackage{amssymb}
\usepackage{amsmath}
\usepackage{amsthm}
\usepackage{graphicx,bm}
\usepackage{epstopdf}
\usepackage{subfigure}
\usepackage{natbib}
\usepackage{epsfig}
\usepackage{amsfonts}
\usepackage{mathrsfs}
\usepackage{CJK}
\usepackage[toc,page,title,titletoc,header]{appendix}
\usepackage{array,longtable}
\usepackage{float}
\usepackage[usenames, dvipsnames, x11names]{xcolor}
\usepackage[colorlinks]{hyperref}
\usepackage[capitalise]{cleveref}
\usepackage{enumerate}
\usepackage{soul}
\usepackage{dsfont}
\hypersetup{
 colorlinks=true,
 citecolor=red,
 linkcolor=blue,
 urlcolor=cyan}
\usepackage{mathtools}
\usepackage{lipsum}

\renewcommand{\H}{\hat{\mathcal{H}}}
\newcommand{\D}{\mathcal{D}}
\newcommand{\U}{\hat{\mathcal{U}}}

\newcommand{\PT}{\mathcal{PT}}

\newcommand{\hc}{\mathrm{H.c.}}

\usepackage{comment}

\begin{document}

\title{Stability via symmetry breaking in interacting driven systems}

\author{Andrew Pocklington}
\email{abpocklington@uchicago.edu}
\affiliation{Department of Physics, University of Chicago, 5640 South Ellis Avenue, Chicago, Illinois 60637, USA }

\affiliation{Pritzker School of Molecular Engineering, University of Chicago, Chicago, IL 60637, USA}

\author{Aashish A. Clerk}
\affiliation{Pritzker School of Molecular Engineering, University of Chicago, Chicago, IL 60637, USA}

\date{\today}

\begin{abstract}
Photonic and bosonic systems subject to incoherent, wide-bandwidth driving cannot typically reach stable finite-density phases using only non-dissipative Hamiltonian nonlinearities; one instead needs nonlinear losses, or a finite pump bandwidth. We describe here a very general mechanism for circumventing this common limit, whereby Hamiltonian interactions can cut-off heating from a Markovian pump, by effectively breaking a symmetry of the unstable, linearized dynamics. We analyze two concrete examples of this mechanism. The first is a new kind of $\PT$ laser, where Hermitian Hamiltonian interactions can move the dynamics between the $\PT$ broken and unbroken phases and thus induce stability. The second uses onsite Kerr or Hubbard type interactions to break the chiral symmetry in a topological photonic lattice, inducing exotic phenomena from topological lasing to the stabilization of Fock states in a topologically protected edge mode.
\end{abstract}


\maketitle 


\textit{Introduction---}
The competition between pumping, loss, and interactions are a ubiquitous feature of both classical and quantum nonlinear dynamics. The often complex interplay between incoherent pumping (which tends to drive a system towards an infinite temperature state) and a nonlinearity that can cut off heating instabilities leads to a rich variety of physics. Examples include classical limit cycles and lasers in the few body regime \cite{Walls1994,Gardiner2004}, classical synchronization \cite{Acebron2005,Dorogovtsev2008}, and driven-dissipative phase transitions in many body systems \cite{Sieberer2013,Diehl2008,Diehl2010_2,Carusotto2013} (e.g.~effective Mott insulator to superfluid transitions in driven bosonic lattice systems \cite{Littlewood2006,Lebreuilly2017, Scarlatella2019,Ma2019}).

\begin{figure}[t]
\centering
\includegraphics[width = 3.25in]{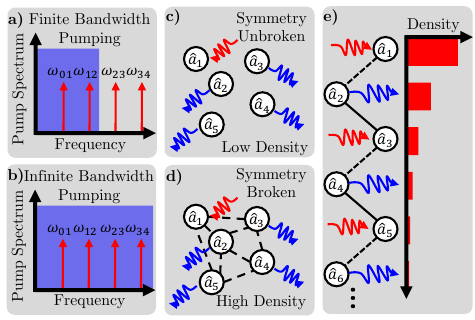}
\caption{a) Standard mechanism for obtaining stable states in interacting bosonic systems subject to a finite-bandwidth, incoherent pump: transition frequencies increase with increasing excitation number, eventually having no overlap with the pump spectrum (blue). b) This mechanism fails for a Markovian, infinite-bandwidth pump. c) A set of orbitals / normal modes, where one mode is guaranteed by some symmetry to experience only pumping (red arrow, $\hat a_1$), while the remaining modes are lossy (blue arrows).
The low-density phase is thus unstable. 
d) A symmetry-breaking interaction forces the modes to hybridize at high photon density, leading to a stable finite-density state. e) By subjecting a topological SSH chain to staggered pumping and loss, symmetry breaking interactions can induce topological lasing, with an exponentially localized steady state photon density.
}
\label{fig:1}
\end{figure}

The simplest form of the above interplay arises when the incoherent pumping has a finite bandwidth. In this case, instability can be prevented with Hamiltonian interactions or nonlinearities that shift system excitation energies with increasing density. This naturally cuts off heating and stabilizes a finite density state, as higher excited states will become non-resonant with the pump, see \cref{fig:1}a.  Such bandwidth-limited pumping has been exploited to realize an effective chemical potential for light, and requires coupling to a non-Markovian (or finite bandwidth) dissipative reservoir \cite{Hafezi2015,Lebreuilly2017,Ma2017_2}. 
Realizing such setups in many-body settings is generally challenging (though see \cite{Ma2019} for a recent experiment). 
If one instead is restricted to completely broadband (Markovian) pumping, then conventional wisdom dictates that Hamiltonian interactions are irrelevant, as pumping is insensitive to system energies [c.f. \cref{fig:1}b]. In this case, one expects that a heating instability can only be prevented by nonlinear loss. Paradigmatic examples here are the van der Pol oscillator \cite{vanderPol1926}, as well standard oscillator/laser gain-saturation mechanisms \cite{Walls1994,Gardiner2004}.


The two above routes would seem to exhaust the set of generic mechanisms for having nonlinearities (either Hamiltonian or dissipative) cut off pumping-induced instabilities. In this Letter, we show that this is surprisingly not the case: there is a third extremely generic mechanism that allows Hamiltonian interactions to 
stabilize systems driven by a broadband (Markovian) pump.
The trick here is not to make use of energy shifts (as the pump is insensitive to energies in this limit), but instead exploit symmetry, see \cref{fig:1}. As we demonstrate in detail, by making use of symmetries that can be effectively broken as a function of excitation density, Hamiltonian interactions combined with incoherent broadband pumping can stabilize finite-density states.   
More concretely, we show that certain systems are sensitive to incoherent pumping only because of symmetries of 
the effective low-density dynamics. By adding a nonlinearity that breaks these symmetries as particle density increases, we obtain a highly effective mechanism for stabilizing finite density states, one that is distinct from more standard approaches.

To illustrate our ideas, we first analyze a quantum version of the well-studied parity-time ($\PT$) symmetric gain-loss dimer \cite{Bender1998,Bender2007}.
While stabilization here is usually achieved via nonlinear dissipation, we show that a purely Hamiltonian approach is also possible: a finite density phase can be stabilized by introducing
non-dissipative quartic nonlinearities. The result is a new kind of $\PT$ laser, e.g. \cite{Feng2014,Peng2014,Ge2016,Hodaei2014,El-Ganainy2018}.
We then turn to more complex many-body bosonic lattice systems subject to Markovian pumping, showing that the combination of a chiral sublattice symmetry and standard on-site Kerr (or Hubbard) interactions allows one to stabilize finite-density states. This mechanism is especially interesting for systems supporting topological edge modes \cite{Schnyder2008}, as it enables a new kind of topological lasing \cite{Hu2021,Secli2021,StJean2017,Parto2018,Zhao2018} and the stabilization of non-classical, topological light \cite{Mittal2018,Blanco-Redondo2018}, e.g. single-photon Fock states in a topological edge mode, \cite{Owens2022}. Surprisingly, this can occur even in the limit of very weak interactions. 

\textit{$\PT$ Symmetry---}
We first consider a parity-time ($\PT$) symmetric dimer: two bosonic modes $\hat a$, $\hat b$ with a tunnel Hamiltonian $\H_0 = J(\hat a^\dagger \hat b + \hat b^\dagger \hat a)$, with $\hat a$ ($\hat b$) experiencing Markovian loss (gain) at a rate $\kappa_a$ ($\kappa_b$). The quantum dynamics of the system's density matrix $\hat{\rho}$ is described by a Lindblad master equation \cite{Lindblad1976,Gorini1976}:
\begin{align}
   \dot{\hat{\rho}} &= -i[ \H_0, \hat \rho] + \kappa_a \D[\hat a] \hat \rho + \kappa_b \D[\hat b^\dagger] \hat \rho.
\end{align}
where $\D[\hat L] \hat \rho \equiv \hat L \hat \rho \hat L^\dagger - \frac{1}{2} \{ \hat L^\dagger \hat L, \hat \rho \}$. 
This system has been studied extensively in both classical \cite{Bender1998,El-Ganainy2018} and quantum \cite{Kepesidis2016} settings, 
and undergoes a $\PT$ breaking transition when $J$ drops below $J_c = \frac{1}{4}(\kappa_a + \kappa_b)$, becoming dynamically unstable when $J \leq \frac{1}{2}\sqrt{\kappa_a \kappa_b}$. This has a simple intuitive explanation: for small $J$, $\PT$ is broken and eigenmodes localize, hence the eigenmode localized on $b$ can become unstable as it primarily experiences gain. When instead $J \gg \kappa_a, \kappa_b$, $\PT$ is unbroken and eigenmodes delocalize. Assuming that $\kappa_a > \kappa_b$, this means both eigenmodes experience net loss, making them dynamically stable.

In the regime $2J<\sqrt{\kappa_a \kappa_b}$, the system is dynamically unstable, and a nonlinearity is needed to restore stability. In a standard $\PT$ laser, this would be done by adding nonlinear dissipation which causes the gain to saturate at a large photon number (see e.g.~\cite{Kepesidis2016,Feng2014}). Here, we propose an alternative stabilization mechanism that only uses a \textit{Hamiltonian} interaction 
which restores $\PT$ as the number of excitations increases. This requires a nonlinear hopping $J \to \tilde J(n)$ where the hopping strength grows monotonically with the total photon number $n$. We consider $\H_{\mathrm{int}} = \frac{J}{2} \frac{\hat n }{n^*}\left( \hat a^\dagger \hat b + \hat b^\dagger \hat a \right)$ where $\hat n \equiv \hat a^\dagger \hat a + \hat b^\dagger \hat b$, and $n^*$ a dimensionless constant setting the
scale of the nonlinearity. 
The total Hamiltonian is thus
\begin{align}
   \H &\equiv \H_0 + \H_{\rm int} =  J\left(1 +  \frac{1}{2}  \frac{\hat n }{n^*} \right) \left( \hat a^\dagger \hat b + \hat b^\dagger \hat a \right).\label{eqn:1}
\end{align}
Note that similar density-dependent tunneling interactions have been studied in a variety of different contexts (e.g.~\cite{Hirsch1989,Maik2013,Hadad2016}).

We study the resulting master equation dynamics using a standard mean-field decomposition \cite{Supplement}. This yields the equations of motion:
\begin{align}
    \dot n_a &= \tilde J c - \kappa_a n_a, \label{eqn:PTEOM1}\ \ \ 
    \dot n_b = - \tilde J c + \kappa_b n_b + \kappa_b, \\
    \dot c &= 2 \tilde J (n_b - n_a) - \frac{\kappa_a - \kappa_b}{2} c,\label{eqn:PTEOM2}
\end{align}
where we have defined the density dependent hopping $\tilde J(n_a,n_b) = J(1 + 1/2 n^*) + J(n_a + n_b)/n^*$, the local densities $n_a \equiv \langle \hat a^\dagger \hat a \rangle, n_b \equiv \langle \hat b^\dagger \hat b \rangle$, and the current $c \equiv i \langle a^\dagger b - b^\dagger a \rangle$. 
The equations can be solved self-consistently, and show that as expected, our Hamiltonian nonlinearity always stabilizes the system in regimes where the linear dynamics is unstable. We give an analytic expression for the steady state density, stability conditions, as well as phase diffusion in the finite density regime in the supplemental materials \cite{Supplement}. We also find that the mean-field description gives a good approximation to exact master equation numerics in the regime of large density \cite{Supplement}.

We stress that there are significant qualitative and quantitative differences between our Hamiltonian stabilization route to $\PT$ lasing 
versus approaches based on standard gain saturation \cite{Walls1994,Kepesidis2016}. In a conventional $\PT$ laser \cite{Feng2014,Hodaei2014}, the gain is naturally cutoff at large photon values, i.e.~the gain rate depends on the number of photons in the gain mode, $\kappa_b \equiv \kappa (1 + \langle \hat n_{b} \rangle/n_{crit} )^{-\nu}$ for some $\nu > 0$ \cite{Kepesidis2016}. Such a \textit{dissipative} nonlinearity can of course stabilize almost \textit{any} kind of pumping-induced instability (i.e.~one is effectively just turning down the pumping strength with increasing density). As such, for this conventional route to stability, the presence or absence of $\PT$ is largely irrelevant. 

In marked contrast to this, the $\PT$ symmetry of the linearized dynamics is critical to stabilization in our nonlinear hopping model. When $\kappa_a = \kappa_b \equiv \kappa$, if one breaks the $\PT$ symmetry by adding a relative detuning $\delta$, $\H \to \H + \delta (\hat a^\dagger \hat a - \hat b^\dagger \hat b)$, then a model using gain saturation to cut off the instability is almost unchanged from the $\delta=0$ case, whereas the nonlinear hopping model immediately becomes unstable when there is a nonzero detuning. This behaviour is independent of the nonlinearity strength, and reflects the fact that with a non-zero $\delta$, it is no longer possible to dynamically tune the system between the $\PT$ broken and unbroken phases by varying density. This stark difference is shown in Fig. S1 of the SM \cite{Supplement}. The strong sensitivity to $\delta$ in our setup could potentially be harnessed for a new kind of non-Hermitian sensing modality \cite{Wiersig2014,Chen2017,Hodaei2017,Zhang2019,Chen2019,McDonald2020_2}. 


\begin{figure}[t]
    \centering
    \includegraphics{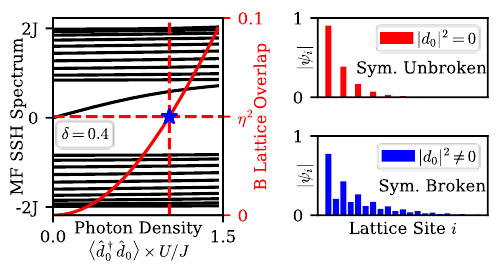}
    \caption{Left. Spectrum (solid black lines) and edge-mode overlap with the B sublattice (solid red) for a mean-field description of our driven-dissipative SSH model, where the effective Hamiltonian is density dependent. Interactions and pumping generate self consistent onsite energies $\Delta_i$, modifying the spectrum and wavefunctions. When the B-lattice overlap matches the gain to loss ratio $\eta^2$, the dynamics achieves stability. This then determines the steady state photon density (dashed red lines). Right: Wavefunction profile of the edge mode in mean-field theory when $U |d_0|^2  = 0$ (upper) and $U |d_0|^2  = 1.4$ (lower).  When the steady state density is $0$ chiral symmetry is unbroken, and the edge-mode wavefunction vanishes on the $B$ sublattice. As the density increases, the chiral symmetry is broken, 
    causing the edge mode to have overlap with both sublattices. In all plots we have taken $N = 21$ sites, and $\delta = 0.4$.
    }
    \label{fig:2}
\end{figure}


\textit{Chiral Symmetry---}
We now consider a more general version of our stabilization mechanism, where the relevant interactions break a purely Hamiltonian symmetry at non-zero densities. We will consider the example of photons hopping on a lattice with a chiral symmetry, and use on-site Kerr nonlinearities (i.e.~Hubbard interactions $U$) to provide the requisite symmetry breaking. Note that this is distinct from previous studies exploring the consequences of squeezed dissipation in chiral bosonic lattices having linear dynamics \cite{Zippilli2015,Ma2017,Yanay2018,Zippilli2021,Pocklington2023}. Our work involves no squeezing, and is focused on {\it nonlinear} dynamics (something not treated in these previous works).

Let $\H_0$ be a quadratic Hamiltonian that has a chiral sublattice symmetry $\U: \U \H_0 \U^\dagger \mapsto -\H_0$. Examples include nearest neighbor tight-binding models on any bipartite lattice. A particularly simple example is a 1D lattice $\H_0 = \sum_i J_i \hat a_i^\dagger \hat a_{i + 1} + \hc$ The chiral symmetry here is $\U: \U \hat a_{i} \U^\dagger \mapsto (-1)^i \hat a_i$.
More generically, such a chiral symmetry defines two sublattices $A,B$. If we have local annihilation operators $\hat a_i \in A$ and $\hat b_i \in B$, then
\begin{align}
    \U \hat a_i \U^\dagger &= \hat a_i, \ \ \ \U \hat b_i \U^\dagger = - \hat b_i.
\end{align}
We consider a setup where we apply very weak incoherent pumping (gain) with a rate $\eta^2 \kappa$ ($\eta \ll 1$) on the $A$ sublattice, and loss with a rate $\kappa$ on the $B$ sublattice. This is described by the Lindblad master equation:
\begin{align}
    \dot{\hat{\rho}} &= -i[\H, \hat \rho] + \kappa \left( \sum_{i} \eta^2 \D[\hat a_{i}^\dagger] + \D[\hat b_i]  \right) \hat \rho \label{eqn:chiral_me}
\end{align}
with $\H = \H_0 + \H_{\rm int}$. 
The hopping Hamiltonian $\H_0$ can be diagonalized as $\H_0 = \sum_\alpha \epsilon_\alpha \hat d_\alpha^\dagger \hat d_\alpha$. By making suitable unitary rotations of the jump operators, we can rewrite the dissipative part of the master equation as
\begin{align}
   \sum_{i} \D[\hat a_{i}^\dagger] &= \frac{1}{4} \sum_\alpha \D[\hat d_\alpha^\dagger + \U \hat d_\alpha^\dagger \U^\dagger], \label{eq:diss_1} \\
   \sum_{i} \D[\hat b_i] &= \frac{1}{4} \sum_\alpha \D[\hat d_\alpha - \U \hat d_\alpha \U^\dagger], \label{eq:diss_2}
\end{align}
using nothing more than the chiral symmetry \cite{Supplement}. Note that $\U \hat d_\alpha \U^\dagger$ is necessarily also an eigenmode annihilation operator with energy $-\epsilon_\alpha$. 

Consider first the case where there are no interactions, and $\H_0$ has a non-zero spectral gap $\Delta$ ($|\epsilon_\alpha| > \Delta > 0$) such that $\kappa \ll \Delta$. We can then approximate $\D[\hat d_\alpha^{(\dagger)} \pm \U \hat d_\alpha^{(\dagger)} \U^\dagger] \sim \D[d_\alpha^{(\dagger)}] + \D[\U \hat d_\alpha^{(\dagger)} \U^\dagger]$. Each mode thus experiences a net decay rate
 $\kappa(1 - \eta^2)/2 \sim \kappa/2$, and the dissipative steady state is simple: each mode will be in a thermal state with average photon number
\begin{align}
    n_{th} &= \frac{\eta^2}{1 - \eta^2} \ll 1.
\end{align}

The physics is more interesting in the case where $\H_0$ has $n > 0$ zero energy eigenmodes $\hat d_{0,i}, \ i = 1,\dots, n$, that are mapped to themselves by the chiral symmetry, $\U \hat d_{0,i} \U^\dagger = \hat d_{0,i}$. \cref{eq:diss_1,eq:diss_2} imply that these modes will \emph{only} experience incoherent pumping and not any loss. These modes are thus unstable regardless of how small $\eta$ is, with a gain rate of $\eta^2 \kappa$. Intuitively, this is a consequence of the zero modes' wavefunctions only having support on the $A$ sublattice. 

We now introduce on-site Kerr (Hubbard) interactions:
\begin{align}
    \H_{\mathrm{int}} &= \frac{U}{2}\sum_i \left( \hat a_i^\dagger \hat a_i^\dagger \hat a_i \hat a_i + \hat b_i^\dagger \hat b_i^\dagger \hat b_i \hat b_i \right).
\end{align} 
This interaction explicitly breaks the chiral sublattice symmetry. Hence, at a mean-field level (where we describe the system with a self-consistent quadratic Hamiltonian), the interaction will cause the unstable, symmetry protected zero modes to hybridize with the stable lossy modes as the dynamics drives the system from low to high density. This hybridization will induce loss into the original unstable modes, leading ultimately to a stable, finite-density state. In what follows, we give a specific example of this mechanism in a lattice with a protected topological edge mode, showing that the mean-field picture sketched here is accurate at large photon density and predicts topological lasing. We further show that the mechanism also works in non-mean field low density regimes, leading to edge states stabilized in non-classical states. 

\begin{figure}[t]
    \centering
    \includegraphics[width = 3.4in]{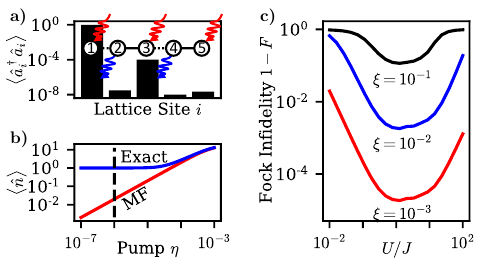}
    \caption{
    a) Exact steady-state photon density for a 5 site driven-dissipative interacting SSH chain (c.f.~\cref{eq:SSH_Ham,eqn:chiral_me}) with $U = \kappa = J$, $\xi = 10^{-2}$, and $\eta = \xi^2$. 
    Despite the global, Markovian pumping in the system, the photonic density is strongly localized at the left edge.
    Inset depicts the first five sites of an SSH chain where solid (dashed) lines represent strong (weak) bonds. Even (odd) numbered sites experience loss shown with blue arrows (gain, red arrows), and each site has a local Kerr nonlinearity.
    b) Steady state total photon number versus the pump to loss ratio $\eta$ obtained from mean field theory (red) and from exact numerics (blue). Exact numerics asymtote to 1 photon in the weak pump limit as it approaches a Fock state in the edge mode for $\eta \lesssim \xi^2$ (dashed vertical line). Here, $U = 0.1, \xi = 10^{-3}, \kappa = J = 1$.
    c) (In)fidelity to a single photon Fock state in the topological edge mode for $\xi = 10^{-1}$ (black), $10^{-2}$ (blue), and $10^{-3}$ (red) versus the nonlinearity $U$. Here, $\kappa = J = 1, \eta = \xi^2$.
    }
    \label{fig:3}
\end{figure}


\textit{Topological Lasing---}
Generating a lasing transition in a photonic lattice where only a topological edge lases is a difficult engineering challenge. Many solutions often require injecting extremely narrowband pumping at the edge-mode frequency \cite{Peano2016,Secli2021}. Our symmetry-breaking stabilization mechanism opens a new pathway for realizing this goal using completely broadband, Markovian pumping. We will consider a concrete example of a $1D$ bosonic SSH chain \cite{Su1979,Su1980}, something that has been realized in a variety of experimental platforms, from superconducting circuits \cite{Kim2021}, micropillar polariton cavities \cite{StJean2017}, photonic cavities \cite{Gong2021}, ring resonators \cite{Leykam2020,Parto2018,Zhao2018}, and optomechanics \cite{Youssefi2022}. With an odd number of sites, this lattice has a single, topologically protected, zero energy edge mode, where the topology is protected by the chiral sublattice symmetry. The noninteracting Hamiltonian is
\begin{align}
    \H_{\mathrm{SSH}} &= -J \sum_{i = 1}^{N-1} (1 + \delta) \hat a_i^\dagger \hat b_{i} + (1 - \delta) \hat b_i^\dagger \hat a_{i + 1}  + \hc \label{eq:SSH_Ham}
\end{align}
where $-1 < \delta < 1$. This system is in a topological regime when $\delta < 0$, and there exists a single zero energy edge mode that lives on only the $A$ sublattice (i.e., $n = 1$). 

Following our general recipe, we apply incoherent pumping to the $A$ sublattice and loss to the $B$ sublattice, and again consider an interaction $ \H_{\mathrm{int}} = \frac{U}{2} \sum_i \hat a_i^\dagger \hat a_i^\dagger \hat a_i \hat a_i + \hat b_i^\dagger \hat b_i^\dagger \hat b_i \hat b_i$. The full master equation is then given by \cref{eqn:chiral_me} with $\H = \H_{\mathrm{SSH}} + \H_{\mathrm{int}}$. This corresponds to a specific example of a driven-dissipative SSH Bose-Hubbard chain, with alternating gain and loss. The semiclassical (Gross–Pitaevskii) equations for the dynamics of $a_i \equiv \langle \hat a_i \rangle$, $b_i \equiv \langle \hat b_i \rangle$ (c.f. Eqs.~(S45)~and~(S46) in \cite{Supplement}) involve self-consistent onsite potentials $\Delta_i$:  
\begin{align}
        &\Delta_i = U |\psi_i|^2 |d_0|^2, \label{eqn:MFSSH_A} \\
    & \sum_{j} \left[ (H_{\mathrm{SSH}})_{ij} + \Delta_i \delta_{ij}\right] \psi_j   = \lambda \psi_i.  \label{eqn:MFSSH_B} 
\end{align}
$d_0$ is the self-consistent edge mode amplitude (which depends on the local onsite energy $\Delta_i$), $\psi_i$ its real-space wavefunction, and $(H_{\mathrm{SSH}})_{ij}$ the single-particle SSH Hamiltonian matrix. 
As can be seen in \cref{fig:2}, as the self-consistent onsite energy $U |d_0|^2 \equiv \sum_i \Delta_i$ is increased from zero (i.e., the photon density increases), the overlap of the self-consistent edge mode on the lossy $B$ sublattice also increases, reflecting the breaking of chiral symmetry and inducing stability in this mode. 
This delocalization leads to a finite-density steady state.  For weak $\kappa$, the corresponding density is found by the value of $U |d_0|^2 $ that makes the spatial overlap of the edge mode on the $B$-lattice exactly $\eta^2$ (the pumping to loss ratio). The breaking of chiral symmetry also causes the edge mode to acquire a nonzero energy [c.f. \cref{fig:2}].  As a result, we obtain a stable limit cycle where the zero mode oscillates at the self-consistent frequency generated by the Kerr nonlinearity \cite{Supplement}.

\textit{Topological Fock States---}
In parameter regimes we are able to test numerically, the semiclassical analysis of \cref{eqn:chiral_me} is accurate if the stabilized photon number is large. However, the stabilization mechanism is not limited to these parameter regimes. Consider our driven SSH chain in the regime where $U \sim J \sim \kappa$, and the dimensionless pumping parameter $\eta$ is chosen so that semiclassics would predict 
a steady state with $\mathcal{O}(1)$ photons in the edge mode. In this regime, we show that it is possible to instead stabilize a single photon Fock state in the edge, where the steady state density matrix can be approximated as $\hat \rho_{ss} \sim \hat d_0^\dagger |0\rangle \langle 0 | \hat d_0 \equiv \hat \rho_1$, with $|0\rangle$ the vacuum and $\hat d_0$ the annihilation operator of the symmetry-protected edge mode, see \cref{fig:3}a. It has been a long-standing challenge to engineer non-classical light in a topological edge mode \cite{Owens2022}, and the proposed stabilization mechanism provides an alternative method.

To understand the origin of this effect, we first define the localization length of the edge mode $\xi \equiv (1 - \delta)/(1 + \delta)$, and write the pumping rate parameter $\eta$ in \cref{eqn:chiral_me} as $\eta \equiv \eta' \xi^2$.  Consider the physics in the strong localization case where $\xi \rightarrow 0$ while $\eta' = \mathcal{O}(1)$ stays fixed.  To leading order in $\xi^2$, we find that the steady state is exactly a Fock state, see \cref{fig:3}c. This is intimately related to the stabilization mechanism: because the linear dynamics are unstable, and the discreteness of the nonlinearity means that it only acts non-trivially when there are at least two photons in the lattice, the quantum dynamics can {\it never} stabilize fewer than a single photon in the steady state \cite{Supplement}, see \cref{fig:3}b. The upshot is that one can stabilize a Fock state to arbitrary accuracy in this regime by choosing $\xi$ appropriately.

Despite the fact that the steady state in this case only has density in a single mode, we stress the fact that the physics here cannot be captured by any simple effective single-mode theory, e.g. single photon pumping and two-photon loss \cite{Supplement}. The stabilization mechanism is completely reliant on the fact that the edge mode is able to weakly hybridize with the stable bulk modes as the photon density increases. Furthermore, our numerically-exact master equation simulations show that our Fock state stabilization mechanism is extremely robust against variations. The nonlinearity can be an order of magnitude weaker than the loss rate and the quadratic Hamiltonian ($U \ll \kappa, J$) with little impact on the final fidelity, see \cref{fig:3}c. This is a further potential advantage of our method: it can generate Fock states with extremely weak nonlinearities.

\textit{Conclusion---}
In this paper, we have both introduced and analyzed a new generic method that allows non-dissipative Hamiltonian interactions and nonlinearities to stabilize finite density phases in bosonic systems subject to broadband, incoherent pumping. The mechanism does not rely on energetics, but instead uses the ability of interactions to effectively break symmetries at finite density. 
Our method is highly adaptable, and represents a generic mechanism for using incoherent, Markovian pumping and Hamiltonian nonlinearities to stabilize finite density phases of matter in bosonic lattices (including non-classical states such as Fock states). While we focused here on chiral symmetry, we expect that our mechanism can also be used to stabilize topological lasing in lattice models protected by different kinds of symmetry.

\textit{Note Added---}
During the final preparation of this work, we became aware of \cite{yang2023}, which identified a related mechanism for selectively populating edge modes in a non-interacting fermionic model. Note that the physics in the fermionic case is very different, as unlike bosons, there is no possibility of dynamical instabilities, hence finite density states of driven setups emerge generically even without interactions or nonlinearity. 

\textit{Acknowledgements---}
This work is supported by the Air Force Office of Scientific Research under Grant No. FA9550-19-1-0362, and was partially supported by the University of Chicago Materials Research Science and Engineering Center, which is funded by the National Science Foundation under Grant No. DMR-1420709. A. C. also acknowledges support from the Simons Foundation through a Simons Investigator Award (Grant No. 669487, A. C.).

\bibliography{ref}

\end{document}


\title{Supplemental Material: Stability via symmetry breaking in interacting driven systems}

\author{Andrew Pocklington}
\affiliation{Department of Physics, University of Chicago, 5640 South Ellis Avenue, Chicago, Illinois 60637, USA }

\affiliation{Pritzker School of Molecular Engineering, University of Chicago, Chicago, IL 60637, USA}

\author{Aashish A. Clerk}
\affiliation{Pritzker School of Molecular Engineering, University of Chicago, Chicago, IL 60637, USA}

\date{\today}

\maketitle

\renewcommand{\theequation}{S\arabic{equation}}
\renewcommand{\thesection}{\Roman{section}}
\renewcommand{\thefigure}{S\arabic{figure}}
\renewcommand{\thetable}{S\arabic{table}}
\renewcommand{\bibnumfmt}[1]{[S#1]}
\renewcommand{\citenumfont}[1]{S#1}


\renewcommand{\theequation}{S\arabic{equation}}
\renewcommand{\thesection}{\Roman{section}}
\renewcommand{\thefigure}{S\arabic{figure}}
\renewcommand{\thetable}{S\arabic{table}}
\renewcommand{\bibnumfmt}[1]{[S#1]}
\renewcommand{\citenumfont}[1]{S#1}

\setcounter{page}{1}
\setcounter{equation}{0}
\setcounter{figure}{0}
\section{$\PT$ Symmetry}

\subsection{Mean Field Solution}

Consider the model proposed in the main text, given by
\begin{align}
    \H &= J\left( 1 + \frac{\hat n}{2 n^*} \right) \left( \hat a^\dagger \hat b + \hat b^\dagger \hat a \right), \label{seqn:pt_ham} \\
    \hat L_1 &= \sqrt{\kappa_a} \hat a, \ \ \ \ \ \ \hat L_2 = \sqrt{\kappa_b} \hat b^\dagger. \label{seqn:pt_jumps}
\end{align}
We can calculate the time evolution of operator averages using the adoint Liouvillian $\partial_t \langle \hat O \rangle = \langle i [\H, \hat O] + \frac{1}{2} \sum_i [\hat L_i^\dagger, \hat O]\hat L_i + \hat L_i^\dagger [\hat O, \hat L_i] \rangle$. We define photon number operators $\hat n_a = \hat a^\dagger \hat a, \hat n_b = \hat b^\dagger \hat b$, and the current operator $\hat c \equiv i(\hat a^\dagger \hat b - \hat b^\dagger \hat a)$. Note that the operator  $\hat a^\dagger \hat b + h.c.$ is conserved by the Hamiltonian dynamics, and so it decays to zero in the steady state. We can calculate that
\begin{align}
    \partial_t \langle \hat n_a \rangle &= -J \left\langle \left( 1 + \frac{\hat n}{2 n^*} \right) \hat c \right\rangle - \kappa_a \langle \hat n_a \rangle, \\
    \partial_t \langle \hat n_b \rangle &= J \left\langle \left( 1 + \frac{\hat n}{2 n^*} \right) \hat c \right\rangle + \kappa_b \langle \hat n_b \rangle + \kappa_b, \\
    \partial_t \langle \hat c \rangle &= -2J\left\langle \left( 1 + \frac{\hat n}{2 n^*} \right) (\hat n_a - \hat n_b) \right\rangle - \frac{\kappa_a - \kappa_b}{2} \langle \hat c \rangle.
\end{align}
At this point, what we have is exact, but the equations do not close on themselves (i.e.~the RHS involves three and four-point averages).  To make progress, we make a Gaussian approximation, and evaluate correlators on the RHS by ignoring higher cumulants (i.e. we assume that Wick's theorem holds).   
This yields
\begin{align}
    \langle \hat n \hat c \rangle &= i\langle \hat a^\dagger \hat a \hat a^\dagger \hat b \rangle + i\langle \hat b^\dagger \hat b \hat a^\dagger \hat b \rangle - i\langle \hat a^\dagger \hat a \hat b^\dagger \hat a \rangle - i\langle \hat b^\dagger \hat b \hat b^\dagger \hat a \rangle \\
    &= i\langle \hat a^\dagger \hat a^\dagger \hat a \hat b \rangle + i\langle \hat b^\dagger \hat a^\dagger \hat b  \hat b \rangle - i\langle \hat a^\dagger \hat b^\dagger \hat a  \hat a  \rangle - i\langle \hat b^\dagger \hat b^\dagger \hat b  \hat a \rangle + \langle \hat c \rangle \\
    &= 2i\langle \hat n_a \rangle \langle \hat a^\dagger \hat b \rangle + 2i\langle \hat n_b \rangle \langle \hat a^\dagger  \hat b \rangle - 2i\langle \hat n_a \rangle \langle \hat b^\dagger   \hat a  \rangle - 2i\langle \hat n_b \rangle \langle \hat b^\dagger   \hat a \rangle + \langle \hat c \rangle \\
    &= \left( 2 \langle \hat n \rangle + 1 \right) \langle \hat c \rangle, \\
    \langle \hat n (\hat n_a - \hat n_b) \rangle &= \langle \hat n_a^2 \rangle - \langle \hat n_b^2 \rangle = 2 \langle \hat n_a \rangle^2 - 2 \langle \hat n_b \rangle ^2 + \langle \hat n_a \rangle - \langle \hat n_b \rangle) \\
    &= \left( 2 \langle \hat n \rangle + 1 \right) \langle \hat n_a - \hat n_b \rangle.
\end{align}
Putting everything together, this gives us the result quoted in the main text: 
\begin{align}
    \partial_t \langle \hat n_a \rangle &= -J \left\langle1 + \frac{\hat n + 1/2}{n^*}  \right\rangle \langle \hat c \rangle - \kappa_a \langle \hat n_a \rangle, \\
    \partial_t \langle \hat n_b \rangle &= J \left\langle 1 + \frac{\hat n + 1/2}{n^*}  \right\rangle \langle \hat c \rangle + \kappa_b \langle  \hat n_b \rangle + \kappa_b, \\
    \partial_t \langle \hat c \rangle &= -2J\left\langle 1 + \frac{\hat n + 1/2}{n^*}  \right\rangle \langle \hat n_a - \hat n_b \rangle - \frac{\kappa_a - \kappa_b}{2} \langle \hat c \rangle.
\end{align}
Note that in this mean field approximation, the interaction acts exactly as expected, where the coupling strength increases linearly with the total number of photons in the cavity. The bare coupling strength is increased by a factor of $\frac{1}{2n^*}$ which are related to the zero point fluctuations coming from the canonical commutation relations.

\subsection{Semiclassical Stability Analysis and Phase Noise}

The mean field analysis above allowed us to derive an equation of motion that corresponded to a density dependent hopping. We can similarly consider a semiclassical equation of motion for the field operators $a \equiv \langle \hat a \rangle, b \equiv \langle \hat b \rangle$ starting from \cref{seqn:pt_ham,seqn:pt_jumps}. This gives
\begin{align}
    \dot{a} &= iJ \left( 1 + \frac{|a|^2 + |b|^2}{2n^*} \right) b + \frac{i J a}{2 n^*}(a^* b + b^* a) - \frac{\kappa_a}{2} a, \\
    \dot{b} &= iJ \left( 1 + \frac{|a|^2 + |b|^2}{2n^*} \right) a + \frac{i J b}{2 n^*}(a^* b + b^* a) + \frac{\kappa_b}{2} b.
\end{align}
We can simplify this significantly by recalling that in the full quantum system, we know that $\langle \hat a^\dagger \hat b + \hat b^\dagger \hat a \rangle$ is always zero in the steady state and conserved by the Hamiltonian. We therefore will neglect the term $(a^* b + b^* a)$ in the equations of motion (which does not affect the fixed point or stability), to get the much simpler dynamical system of equations
\begin{align}
    \partial_t a &= -i J \left( 1 + \frac{|a|^2 + |b|^2}{2n^*} \right) b - \frac{\kappa_a}{2} a, \label{seqn:pt_dimer1} \\
    \partial_t b &= -i J \left( 1 + \frac{|a|^2 + |b|^2}{2n^*} \right) a + \frac{\kappa_b}{2} b. \label{seqn:pt_dimer2}
\end{align}
Note that in this form, the nonlinearity acts identically to the mean field case, where the coupling $J$ scales with density.
We can calculate the steady state populations analytically, which gives two solutions:
\begin{align}
    \frac{n}{2n^*} &= 0, \ \ \ \ \ \ \frac{n}{2n^*} = \frac{\sqrt{\kappa_a \kappa_b}}{2J} - 1.
\end{align}
It is trivial to observe that below threshold, when $\sqrt{\kappa_a \kappa_b} < 2 J$, the first solution is dynamically stable, as in this case the nonlinearity does not affect the linear stability analysis. We will next show that in the lasing regime $\sqrt{\kappa_a \kappa_b} > 2 J$, the other solution is dynamically stable.

It is worth noting that the absolute phase of the modes is unconstrained by the fixed point equations, reflecting the $U(1)$ symmetry inherent in the problem. In a real laser, noise would cause this free phase to diffuse, which we can model using the Langevin equations
\begin{align}
    \partial_t a &= -i J \left( 1 + \frac{|a|^2 + |b|^2}{2n^*} \right) b - \frac{\kappa_a}{2} a + \sqrt{\kappa_a} \xi_a, \\
    \partial_t b &= -i J \left( 1 + \frac{|a|^2 + |b|^2}{2n^*} \right) a + \frac{\kappa_b}{2} b + \sqrt{\kappa_b} \xi_b,
\end{align}
where $\langle \xi_{a,b}(t) \xi_{a,b}^*(t')\rangle = \delta(t - t')$ are complex white noise operators. We will expand the equations of motion about the fixed point and define
\begin{align}
a &= (\rho_a + \delta \rho_a) e^{i \phi}, \ \ \ \ \rho_a^2 =  2n^* \frac{\kappa_b}{\kappa_a + \kappa_b} \left( \frac{\sqrt{\kappa_a \kappa_b}}{2J} - 1 \right), \\
b &= i(\rho_b + \delta \rho_b) e^{i (\phi + \delta \phi)}, \ \ \ \ \rho_b^2 =  2n^* \frac{\kappa_a}{\kappa_a + \kappa_b} \left( \frac{\sqrt{\kappa_a \kappa_b}}{2J} - 1 \right).
\end{align}
Here, $\delta \rho_{a,b}$ are the fluctuating densities of the $a,b$ modes, respectively, which we will show are constrained to be zero in the steady state without any noise. The relative phase $\delta \phi = 0$ as well in the steady state, with $\phi$ the free absolute phase. Making the approximation that $\delta \rho_{a,b}/\rho_{a,b}, \delta \phi \ll 1$, we can linearize the equations of motion about the stable fixed point to get:
\begin{align}
    e^{-i \phi} \dot a &= \dot{\delta \rho_a} + i \dot \phi (\rho_a + \delta \rho_a) \nonumber \\
    &= J \left( 1 + \frac{(\rho_a + \delta \rho_a)^2 + (\rho_b + \delta \rho_b)^2}{2n^*} \right) (\rho_b + \delta \rho_b) e^{i \delta \phi} - \frac{\kappa_a}{2} (\rho_a + \delta \rho_a) + \sqrt{\kappa_a} e^{-i \phi} \xi_a, \label{eqn:eom_noise_a}\\
    -i e^{-i (\phi + \delta \phi)} \dot b &= \dot{\delta \rho_b} + i (\dot \phi + \dot{\delta \phi})(\rho_b + \delta \rho_a) \nonumber \\
    &= -J \left( 1 + \frac{(\rho_a + \delta \rho_b)^2 + (\rho_b + \delta \rho_b)^2}{2n^*} \right) (\rho_a + \delta \rho_a) e^{-i \delta \phi} + \frac{\kappa_b}{2} (\rho_b + \delta \rho_b) + \sqrt{\kappa_b} e^{-i (\phi + \delta \phi)} \xi_b \label{eqn:eom_noise_b}.
\end{align}
From here, it is simple to define the new (real valued) white noise variables $\xi_a' = \mathrm{Im} \ e^{-i \phi} \xi_a, \ \xi_a'' = \mathrm{Re} \ e^{-i \phi} \xi_a$ and $\xi_b' = \mathrm{Im} \ e^{-i (\phi + \delta \phi)} \xi_b, \ \xi_b'' = \mathrm{Re} \ e^{-i (\phi + \delta \phi)} \xi_b$. By taking the real part of \cref{eqn:eom_noise_a,eqn:eom_noise_b}, we can show that, to linear order, the densities will be stable around the fixed point by looking at the linearized dynamcics:
\begin{align}
    \dot{\delta \rho_a} &= \left( \frac{\sqrt{\kappa_a \kappa_b}}{2} + \frac{\kappa_a}{\kappa_a + \kappa_b} \left( \sqrt{\kappa_a \kappa_b} - 2 J \right) \right) \delta \rho_b + \left( \frac{\sqrt{\kappa_a \kappa_b}}{\kappa_a + \kappa_b} \left( \sqrt{\kappa_a \kappa_b} - 2 J  \right) - \frac{\kappa_a}{2} \right) \delta \rho_a + \sqrt{\kappa_a} \xi_a'', \\
    \dot{\delta \rho_b} &= -\left( \frac{\sqrt{\kappa_a \kappa_b}}{2} + \frac{\kappa_b}{\kappa_a + \kappa_b} \left( \sqrt{\kappa_a \kappa_b} - 2 J \right) \right) \delta \rho_a - \left( \frac{\sqrt{\kappa_a \kappa_b}}{\kappa_a + \kappa_b} \left( \sqrt{\kappa_a \kappa_b} - 2 J  \right) - \frac{\kappa_b}{2} \right) \delta \rho_b + \sqrt{\kappa_b} \xi_b''.
\end{align}
This corresponds to a dynamical matrix whose eigenvalues are
\begin{align}
    -\frac{\kappa_a^2 - \kappa_b^2}{4 \kappa_a \kappa_b} \left(1 \pm \sqrt{1 + \frac{16\sqrt{\kappa_a \kappa_b}}{(\kappa_a - \kappa_b)^2}(2J - \sqrt{\kappa_a \kappa_b})}  \right),
\end{align}
which are negative definite if and only if $2J - \sqrt{\kappa_a \kappa_b} < 0$ as desired. Similarly, we can take the imaginary part of \cref{eqn:eom_noise_a,eqn:eom_noise_b} to show that:
\begin{align}
    \dot{\phi} &= \frac{\kappa_a}{2} \delta \phi + \frac{\sqrt{\kappa_a}}{\rho_a} \xi_a', \\
    \dot{\delta \phi} &= -\frac{\kappa_a - \kappa_b}{2} \delta \phi + \frac{\sqrt{\kappa_b}}{\rho_b} \xi_b' - \frac{\sqrt{\kappa_a}}{\rho_a} \xi_a'.
\end{align}
Note that this also shows linear stability for $\delta \phi$ since $\kappa_a - \kappa_b > 0$.
Now, we can go to Fourier space and solve these explicitly as in terms of the unitless susceptibility $\chi(\omega) \equiv \left( \frac{\kappa_a/2}{i \omega + (\kappa_a - \kappa_b)/2} \right)$ and the pump/loss ratio $r \equiv \kappa_b/\kappa_a$:
\begin{align}
    \delta \phi(\omega) &= \frac{\sqrt{\kappa_a}}{\rho_a} \frac{\chi(\omega)}{\kappa_a/2} \left(r \xi_b' -  \xi_a' \right), \\
    \implies \phi(\omega) &=  \frac{\sqrt{\kappa_a}}{\rho_a} \frac{1}{i \omega} \left[ r \chi(\omega) \xi_b' + (1 - \chi(\omega)) \xi_a' \right].
\end{align}
From here, we can finally extract the diffusion constant by considering the long time behavior of the correlation function $\langle (\phi(t) - \phi(0))^2 \rangle$, where $\langle \cdot \rangle$ represents a noise average.
\begin{align}
    \langle (\phi(t) - \phi(0))^2 \rangle &= \int_{-\infty}^\infty \left( e^{i \omega t} - 1 \right) \left( e^{i \omega' t} - 1 \right) \langle \phi(\omega) \phi(\omega') \rangle \dd \omega \dd \omega'  \nonumber \\
    &= \frac{2 \pi \kappa_a}{\rho_a^2} \int_{-\infty}^\infty \frac{\ \left( e^{i \omega t} - 1 \right) \left( e^{i \omega' t} - 1 \right)}{-\omega \omega'} \left[ r^2 \chi(\omega) \chi(\omega') \delta(\omega + \omega') + (1- \chi(\omega) )(1-\chi(\omega')) \delta(\omega + \omega') \right] \dd \omega \dd \omega'\nonumber \\
    &= \frac{4 \pi \kappa_a}{\rho_a^2} \int_{-\infty}^\infty \frac{1 - \cos(\omega t)}{\omega^2} \left[ r^2 |\chi(\omega)|^2 + |1- \chi(\omega)|^2 \right] \dd \omega \nonumber \\
    &= \frac{8 \pi^2 \kappa_a}{\rho_a^2} \frac{r^2}{(1-r)^2} |t|  \ \ \ \ \text{as} \ \ \ \  t \to \infty.
\end{align}
This gives us the standard Schawlow-Townes phase diffusion constant, modified by the pump/loss ratio. Note that when $r \ll 1$, then 
\begin{align}
    \frac{\kappa_a}{\rho_a^2} \frac{r^2}{(1-r)^2} &\approx \frac{\kappa_a}{\rho_a^2} r^2 =  \frac{\kappa_b}{\rho_b^2},
\end{align}
which tells us that when $r \ll 1$ (and the number of photons in the $b$ mode is much larger than the $a$ mode), then the phase diffusion of the $a$ mode is essentially locked to the Schawlow-Townes value of the $b$ mode.

\subsection{Gain Saturation}

The more standard interaction to stabilize a $\PT$ dimer that occurs often in experiments is gain saturation, where the gain rate is dependent upon the number of photons in the gain mode. This can be described using the semiclassical equations of motion
\begin{align}
    \partial_t a &= -i J  b - \frac{\kappa_a}{2} a, \label{seqn:gain_sat1} \\
    \partial_t b &= -i J  a + \frac{\kappa_b}{2} \frac{1}{1 + |b|^2/n^*} b. \label{seqn:gain_sat2}
\end{align}
We can similarly calculate that there are two solutions: either $a = b = 0$, or
\begin{align}
    \frac{|a|^2}{n^*} &= \frac{\kappa_a \kappa_b - 4 J^2}{\kappa_a^2}, \ \ \ \ \frac{|b|^2}{n^*} = \frac{\kappa_a \kappa_b - 4 J^2}{4 J^2}.
\end{align}
\begin{figure}
    \centering
    \includegraphics{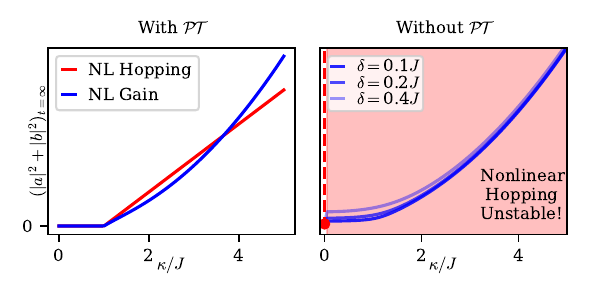}
    \caption{On the left panel, we can see that the steady state photon number undergoes a second order transition at $\kappa = J$ for both the nonlinear hopping model [c.f. \cref{eqn:nonlinear_hop_detuning_a,eqn:nonlinear_hop_detuning_b} with $\delta = 0$] as well as the gain saturation model [c.f. \cref{eqn:gain_sat_delta_a,eqn:gain_sat_delta_b}, $\delta = 0$]. On the right, when we add a detuning and break the single particle $\PT$ symmetry, the gain saturation model remains stable with a steady state photon density that is nearly unchanged (blue curves). On the other hand, the nonlinear hopping model is immediately unstable when $|\delta| > 0$ (red shading) unless $\kappa = 0$ (red dot). In all plots, $n* = 5$.
    }
    \label{fig:S6}
\end{figure}
At this point, we remark that in this gain saturation model, the $\PT$ symmetry breaking generates the instability at the linear level, but is not critical to the actual stabilization mechanism. For example, consider the limit of \cref{seqn:gain_sat1,seqn:gain_sat2} when $\kappa_a = \kappa_b \equiv \kappa$. We can now add a small detuning $\delta$ between the two modes, explicitly breaking the $\PT$ symmetry of the linear dynamics. This is given by the equations of motion:
\begin{align}
    \partial_t a &= -i J  b - i \delta a - \frac{\kappa}{2} a, \label{eqn:gain_sat_delta_a}\\
    \partial_t b &= -i J  a + i \delta b + \frac{\kappa}{2} \frac{1}{1 + |b|^2/n^*} b. \label{eqn:gain_sat_delta_b}
\end{align}
Defining the dynamical matrix to be 
\begin{align}
    D(|b|) &= \left( \begin{array}{cc}
        -i \delta - \frac{\kappa}{2} & -i J \\
        -i J & i \delta + \frac{\kappa}{2(1 + |b|^2/n^*)}
    \end{array} \right),
\end{align}
then whenever $|J|, \kappa > 0$, there exists a value of $|b|$ such that the real part of the spectrum is negative definite. This can be observed by noting that when $|b| \to \infty$ there is no gain at all in the system, only loss, and so the real eigenvalues must be negative semi-definite. However, for there to be an eigenmode that is purely imaginary, then it would have to be completely localized to the $a$ mode, which is impossible if $J \neq 0$. This tells us that the magnitude in the $b$ mode can never grow in an unbounded way, as it will always eventually be cutoff. Thus, the system will eventually stabilize to a fixed average density, even though the $\PT$ symmetry is no longer present.

This is very different from the density-dependent hopping dimer in \cref{seqn:pt_dimer1,seqn:pt_dimer2}. Adding a detuning will again explicitly break the $\PT$ symmetry, giving the equations of motion
\begin{align}
    \partial_t a &= -i J \left( 1 + \frac{|a|^2 + |b|^2}{n^*} \right) b - i \delta a - \frac{\kappa}{2} a, \label{eqn:nonlinear_hop_detuning_a} \\
    \partial_t b &= -i J \left( 1 + \frac{|a|^2 + |b|^2}{n^*} \right) a + i \delta b+ \frac{\kappa}{2} b. \label{eqn:nonlinear_hop_detuning_b}
\end{align}
Defining the linear dynamical matrix in the analogous way, we arrive at
\begin{align}
    D(n) &= \left( \begin{array}{cc}
        -i \delta - \frac{\kappa}{2} & -i J(1 + n/n^*) \\
        -i J(1 + n/n^*) & i \delta + \frac{\kappa}{2}
    \end{array} \right), \\
    \lambda(n) &= \pm \frac{1}{2} \sqrt{(2 \delta - i \kappa)^2 - 4 J^2(1 + n/n^*)^2},
\end{align}
with $\lambda(n)$ the complex eigenvalues. Note that one of these always has a positive real part, regardless of the density $n$, meaning it is always unstable. This reflects the fact that the $\PT$ symmetry was crucial in the stabilization mechanism, as is highlighted in the main text, and shown in \cref{fig:S6}.

\subsection{Beyond Mean Field}

\begin{figure}[h]
    \centering
    \includegraphics[width = 4in]{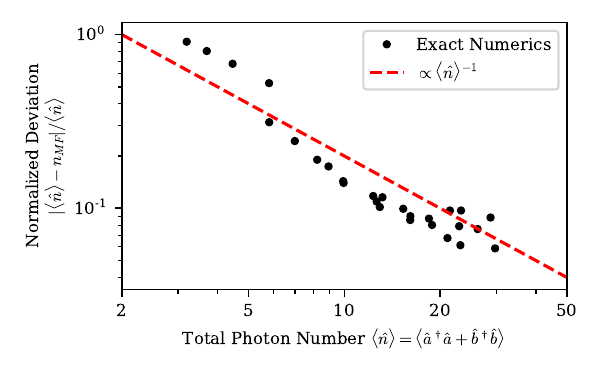}
    \caption{This shows the difference in expectation value of the total photon number between exact numerics and semiclassics, plotted against the total photon number. The dashed red line shows a $1/\langle \hat n \rangle$ scaling. }
    \label{fig:S1}
\end{figure}

Exact diagonalization in systems of up to order 30 steady state photons between the two modes shows that the mean field theory accurately captures the steady state photon density up to an error decaying inversely with the steady state photon number.
This can be observed in \cref{fig:S1}.

\section{Chiral Symmetry}

\subsection{SSH Limit Cycle Solution}

\begin{figure}[h]
    \centering
    \includegraphics[width = 4in]{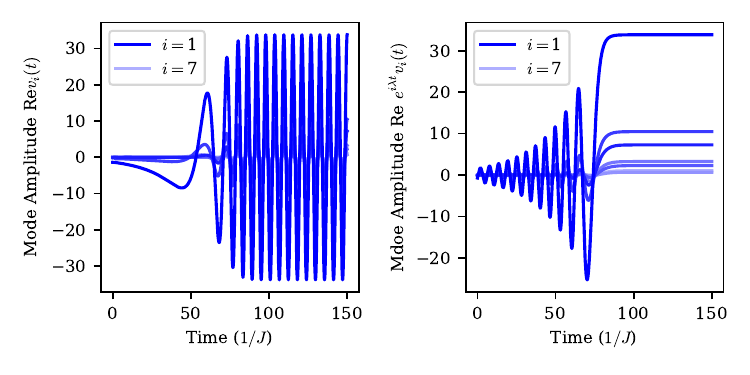}
    \caption{This shows the time trace of the dynamical equation of motion \cref{seqn:2} using $J = 1, U = 0.001, \delta = 0.65, \kappa = 1, \eta = 0.2$. The left plot shows the real part of the amplitude of the dynamical vector $\vec v$, which begins oscillating periodiclaly time in the long time limit. The right shows the same, but in the frame rotating at the critical frequency $\lambda$ defined in \cref{seqn:3}.}
    \label{fig:S4}
\end{figure}

We can identify the limit cycle of the semiclassical equations of motion in the SSH chain as the long term solution to the dynamical equations
\begin{align}
    \dot a_i &= -iJ(1 + \delta) b_{i} - i(1 - \delta) b_{i - 1} - iU|a_i|^2 a_i + \frac{\kappa \eta^2}{2} a_i, \\
    \dot b_i &= -iJ(1 + \delta) a_{i} -i(1 - \delta) a_{i + 1} - iU|b_i|^2 b_i -  \frac{\kappa}{2} b_i,
\end{align}
where there are $N$ different real-space $a$ site amplitudes ($a_1$ through $a_N$) and $N-1$ different real-space $b$ site amplitudes ($b_1$ through $b_{N-1}$). (To make the above consistent with the boundary conditions, simply define $b_0 = b_N \equiv 0$ to get the correct equations of motion for $a_1$ and $a_N$). 

We can define the vector $\vec v = (a_1, b_1, \dots, b_{N-1}, a_N)^T$, and similarly the dynamical matrix $D(\vec v)$ such that
\begin{align}
    \dot{\vec{v}} &= D(\vec v) \vec v, \label{seqn:2}
\end{align}
where the diagonal elements of $D$ include the nonlinearity 
\begin{align}
    D_{ii} = -iU|v_i|^2 + (1 - (-1)^i) \frac{\kappa \eta^2}{4} - (1 + (-1)^i) \frac{\kappa}{4},
\end{align}
and the off diagonal elements are just the SSH coupling. 
As in the main text, we can define the self-consistent edge mode $\vec d_0$, which satisfies the self-consistent eigenvalue equation
\begin{align}
    D(\vec d_0) \vec d_0 = i \lambda \vec d_0, \ \ \ \ \lambda \in  \R.\label{seqn:3}
\end{align}
From here, we can observe that the limit cycle solution is then simply
\begin{align}
    \vec v(t \to \infty) &= e^{i \lambda t} \vec d_0,
\end{align}
since $D(\vec v) = D(e^{i \phi} \vec v) \ \  \forall \phi \in \R$. Numerically, we find that this solution is unique and dynamically stable, as is shown in \cref{fig:S4}. 

\subsection{Gain and Loss in Mode Basis}

Recall that we can write a generic chiral symmetric Hamiltonian in the form 
\begin{align}
    \H &= \sum_{i \in  A, j \in B} H_{ij} \hat a_i^\dagger \hat a_j + \hc \equiv \sum_\alpha \epsilon_\alpha \hat d_\alpha^\dagger \hat d_\alpha,
\end{align}
where $A,B$ are the sublattices. Then we can define the chiral symmetry operator $\U\hat a_i \U^\dagger = \hat a_i, \U\hat b_i \U^\dagger = -\hat b_i $. Let's define the unitary transformation from real space operators to Hamiltonian eigenmodes as
\begin{align}
    \hat d_\alpha &\equiv \sum_i \psi^A_{\alpha}[i] \hat a_i + \psi^B_{\alpha}[i] \hat b_i.
\end{align}
The chiral symmetry tells us that for every mode $\hat d_\alpha$, there is another mode $\hat d_{-\alpha}$ that is also a Hamiltonian eigenmode with opposite energy. It's expansion in real space then must be defined by
\begin{align}
    \hat d_{-\alpha} &\equiv \sum_i \psi^A_{\alpha}[i] \hat a_i - \psi^B_{\alpha}[i] \hat b_i.
\end{align}
We can calculate the inverse of these maps as
\begin{align}
    \hat a_i &= \sum_{\alpha > 0} \psi^A_{\alpha}[i]^* (\hat d_\alpha + \hat d_{-\alpha}), \\
    \hat b_i &= \sum_{\alpha > 0} \psi^B_{\alpha}[i]^* (\hat d_\alpha - \hat d_{-\alpha}),
\end{align}
from unitarity. We assume that we have dissipation in the form of
\begin{align}
    \L &= \eta^2 \kappa \sum_i \D[\hat a_i^\dagger] + \kappa \sum_i \D[\hat b_i].
\end{align}
Recall that the Liouvillian is unchanged under a unitary transformation of the jump operators - i.e 
\begin{align}
    \sum_i \D[\hat L_i] &= \sum_j \D \left[ \sum_i U_{ji} \hat L_i \right],
\end{align}
where $U_{ij}$ is a unitary matrix. Therefore, we can observe the following identites:
\begin{align}
    \eta^2 \kappa \sum_i \D[\hat a_i^\dagger] + \kappa \sum_i \D[\hat b_i]  &= 
    \eta^2 \kappa \sum_i \D \left[ \sum_{\alpha > 0} \psi^A_{\alpha}[i] (\hat d_\alpha^\dagger + \hat d_{-\alpha}^\dagger) \right] + \kappa \sum_i \D\left[ \sum_{\alpha > 0} \psi^B_{\alpha}[i]^* (\hat d_\alpha - \hat d_{-\alpha})  \right]  \\
    &= \frac{\kappa}{2} \sum_{\alpha > 0} \eta^2 \D \left[ \hat d_\alpha^\dagger + \hat d_{-\alpha}^\dagger \right] +  \D\left[ \hat d_\alpha - \hat d_{-\alpha}  \right] \\
    &= \frac{\kappa}{4} \sum_{\alpha} \eta^2 \D \left[ \hat d_\alpha^\dagger + \U \hat d_{\alpha}^\dagger \U \right] + \D\left[ \hat d_\alpha - \U \hat d_{\alpha} \U  \right],
\end{align}
recovering the result in the main text.

\subsection{Effective Nonlinear Hopping}
Let's begin by considering the weakly interacting regime, where we expect thresholdless lasing of the topological edge mode. In the main text, we gave an intuitive picture that the edge mode (localized to the $A$ sublattice) begins to overlap strongly with the $B$ sublattice, the instability is cut off, generating a lasing regime. 

This real space picture is useful for the MF treatment. Alternatively, we could ask what the interaction looks like in the mode picture. As before, let's define quadratic SSH Hamiltonian (but this time not distinguish between sublattices)
\begin{align}
    \H_{\mathrm{SSH}} &= -J \sum_{i = 1}^{N-1} (1 + (-1)^i \delta) \hat a_i^\dagger \hat a_{i + 1} + \hc \equiv \sum_\alpha  \epsilon_\alpha \hat d_\alpha^\dagger \hat d_\alpha, \\ \label{seqn:SSH_HAM}
    \hat d_\alpha &= \sum_i \psi_\alpha[i] \hat a_i,
\end{align}
where we have defined the mode energy $\epsilon_\alpha$ for the $\alpha$ mode, and the real space wavefunction profile $\psi_\alpha[i]$. Again taking the chiral symmetry to be $\U$, it will be useful to define
\begin{align}
    \hat d_{-\alpha} \equiv \U \hat d_\alpha \U^\dagger \implies \epsilon_{-\alpha} = - \epsilon_\alpha,
\end{align}
where we will take $N$ to be odd so there is a singular zero mode, and let $\hat d_{0} = \U \hat d_0 \U^\dagger$. Thus, the quadratic master equation becomes
\begin{align}
    \partial_t \hat \rho &= -i \sum_\alpha \left[ \epsilon_\alpha \hat d_\alpha^\dagger \hat d_\alpha , \hat \rho\right]  + \frac{\kappa}{4} \left( \D[\hat d_\alpha - \hat d_{-\alpha}] + \eta^2 \D[\hat d_\alpha^\dagger + \hat d_{-\alpha}^\dagger]
    \right) \hat \rho.
\end{align}
Now, we want to expand the interaction term in this mode basis:
\begin{align}
    \H_{\mathrm{int}} = \frac{U}{2} \sum_i \hat a_i^\dagger \hat a_i^\dagger \hat a_i \hat a_i &\sim U \sum_i \langle \hat a_i^\dagger \hat a_i \rangle \hat a_i^\dagger \hat a_i = U \sum_{i,\alpha, \beta} \langle \hat a_i^\dagger \hat a_i \rangle \psi_{\alpha}[i] \psi_{\beta}[i] \hat d_\alpha^\dagger \hat d_\beta.
\end{align}
Again, letting $\eta \ll 1$, then $\langle \hat a_i^\dagger \hat a_i \rangle \sim |\psi_0[i]|^2 \langle \hat d_0^\dagger \hat d_0 \rangle$. Next, since $\eta \ll 1$, in the steady state the bulk modes are nearly in vacuum, and so $\hat d_\alpha \hat \rho_{ss} \sim 0 \ \forall \alpha \neq 0$. Therefore, the only relevant terms in the interaction are of the form
\begin{align}
    \H_{\mathrm{int}} & \sim U \langle \hat d_0^\dagger \hat d_0 \rangle \sum_{i,\alpha}  \psi_{\alpha}^*[i] |\psi_{0}[i]|^2 \psi_0[i] \hat d_\alpha^\dagger \hat d_0,
\end{align}
and so this again reduces to just density dependent hopping, this time instead of between modes in the dimer, it is between the edge modes and the lossy bulk modes of the SSH chain. 

\subsection{Topological Fock States}

\begin{figure}
    \centering
    \includegraphics[width = 4.5in]{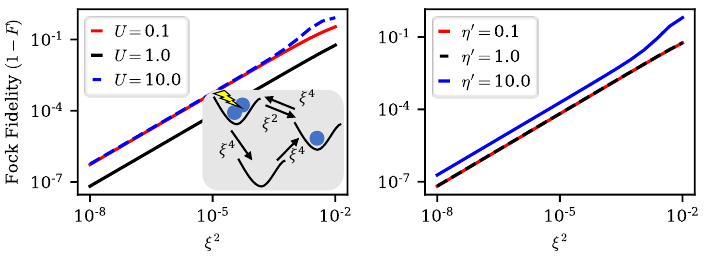}
    \caption{Scaling of the Fock fidelity versus the unitless hopping asymmetry $\xi$, which is also related to the localization length, for various values of $U$ (left) or $\eta'$ (right). Note the predicted quadratic scaling, as well as the strong robustness to parameters. Here, $\eta \equiv \eta' \xi^2, U = J = \kappa = 1$ unless otherwise specified in the plot. Inset: Schematic depicting the Fock stabilization mechanism and scaling properties.}
    \label{fig:S2}
\end{figure}

As observed in the main text, the semiclassics breaks down when it predicts $\mathcal{O}(1)$ photons in the steady state. We will show that the steady state should instead be a topological Fock state, up to corrections that are perturbative in the localization length.

We will define the localization length of the edge mode \begin{align}
    \xi &\equiv \left( \frac{1 + \delta}{1 - \delta} \right),    
\end{align}
and reparameterize $\eta = \eta' \xi^2$, where we will be taking $\xi$ to be a small value and $\eta'$ a dimensionless $\mathcal{O}(1)$ parameter.

The master equation is then
\begin{align}
    \partial_t \hat \rho &= -i \sum_\alpha \left[ \epsilon_\alpha \hat d_\alpha^\dagger \hat d_\alpha + \H_{\mathrm{int}} , \hat \rho\right]  + \frac{\kappa}{4} \left( \D[\hat d_\alpha - \hat d_{-\alpha}] + \eta'^2 \xi^4 \D[\hat d_\alpha^\dagger + \hat d_{-\alpha}^\dagger]
    \right) \hat \rho.
\end{align}
Now, let's first consider the scenario when $\eta' = 0$. In this case, we have only loss, and there are exactly two degenerate steady states. These are the vacuum $\hat \rho_0 \equiv |0\rangle \langle 0|$ and $\hat \rho_1 \equiv \hat d_0^\dagger |0\rangle \langle 0| \hat d_0$ the singly excited edge mode Fock state, which is a steady state because it sees no loss. We cannot, however, add more excitations because the interaction tells us these are no longer eigenmodes of the Hamiltonian. 

Now, let's turn on $\eta'$. To zeroth order, we assume that the perturbation does nothing except within the degenerate steady space manifold. Limiting ourselves to just these two pure states, all the gain wants to do it pump photons from the vacuum into the single Fock state $\hat \rho_1$, which becomes now the non-degenerate steady state.

\begin{figure}[t]
    \centering
    \includegraphics[width = 4in]{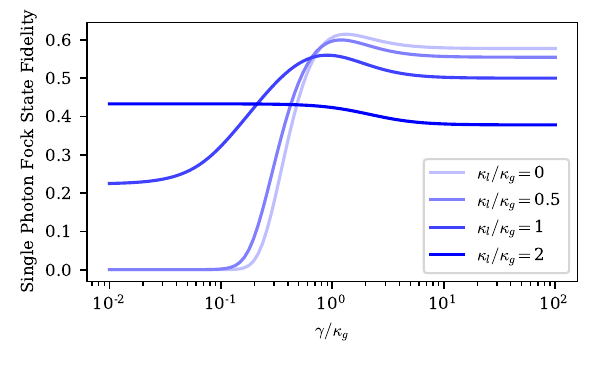}
    \caption{This shows the fidelity of the steady state of the master equation in \cref{seqn:1} as a function of $\gamma/\kappa_g$ for varying values of $\kappa_l/\kappa_g$. Note the optimal value of $\kappa_l = 0$, and $\gamma \sim \kappa_g$, which gives a fidelity of around 0.6.}
    \label{fig:S3}
\end{figure}

To next order, we need to consider the possibility that the pumping can push $\hat \rho_1$ into the doubly excited subspace. However, because $\hat \rho_2 \equiv \hat d_0^\dagger \hat \rho_1 \hat d_0$ does not commute with the Hamiltonian, it has a finite lifetime. Now, we will calculate the different decay rates and channels to leave the doubly excited manifold. Before doing so, it is critical to observe that, by making the localization length small, we make it difficult for the edge mode to tunnel into any other mode:
\begin{align}
    \psi_0[i] &= \sqrt{1 - \xi^2} \sin(\pi i/2) \xi^{(i-1)/2}, \\
    \sum_{\alpha \neq 0} |\langle 0 | \hat d_\alpha \hat d_0 \H \hat d_0^\dagger \hat d_0^\dagger | 0 \rangle| &= \frac{U}{2} \sum_{\alpha \neq 0,i} |\psi_\alpha[i] \psi_0[i]^3| \\
    &= \frac{U}{2} | \psi_0[0]^3| \sum_{\alpha \neq 0} |\psi_\alpha[0]|  + \mathcal{O}(\xi^3) = \mathcal{O}( \xi).
\end{align}
Where the key observation is that the overlap of the edge mode on the first lattice site is $|\psi_0[1]|^2 = 1 - \xi^2 $, and so by the unitarity of the transformation, $\sum_{\alpha \neq 0} |\psi_\alpha[1]|^2 = \xi^2$, and so $\sum_{\alpha \neq 0} |\psi_\alpha[0]| = \mathcal{O}(\xi)$. Now, we can begin to do power counting in powers of $\xi$. For the Liouvillian to map us from  $\hat \rho_2  \to \hat \rho_1$, the non-linearity must hit on both sides of the density matrix to convert the zero mode into another mode [making it $\mathcal{O}(\xi^2)$], and then the loss will decay these particles away [which is $\mathcal{O}(1)$] to give us an overall $\mathcal{O}(\xi^2)$ process. Next, we can ask how we get from the doubly excited state $\hat \rho_2  \to \hat \rho_0$ the vacuum. For this, the non-linearity must hit on both sides twice [making it $\mathcal{O}(\xi^4)$], and then all of these particles decay [again $\mathcal{O}(1)$] giving an overall decay rate that is $\mathcal{O}(\xi^4)$. From here, letting $\rho_i$ to denote the probability to be in state $\hat \rho_i$, we can define the rate equations (just keeping power of $\xi$)
\begin{align}
    \dot \rho_0 &= -\xi^4 \rho_0 + \xi^4 \rho_2, \\
    \dot \rho_1 &= \xi^4 \rho_0 + \xi^2 \rho_2 - \xi^4 \rho_1, \\
    \dot \rho_2 &= \xi^4 \rho_1 - (\xi^4 + \xi^2) \rho_2.
\end{align}
These three equations sum to zero (as expected by the conservation of probability). Solving them gives
\begin{align}
    \rho_1 &= \frac{\xi^2 + \xi^4}{\xi^2 + 3 \xi^4} = 1 - 2 \xi^2 + O(\xi^4), \\
    \rho_0 &= \rho_2 = \frac{\xi^4}{\xi^2 + 3 \xi^4} = \xi^2 + O(\xi^4).
\end{align}
This somewhat heuristic argument is sufficient to perfectly capture the scaling with $\xi$ [c.f. \cref{fig:S3}].

We now wish to show that this cannot be reduced to any simple effective, Markovian, single-mode dynamics. I.e, one cannot use any combination of single-photon pumping, single- or two-photon loss, or any number-conserving Hamiltonian non-linearity to stabilize a single photon Fock state. Consider the single-mode quantum master equation
\begin{align}
    \dot{\hat{\rho}} &= -i\left[ \Delta \hat n + \frac{U}{2} \hat n^2, \hat \rho \right] + \kappa_g \D[\hat a^\dagger] \hat \rho + \kappa_l \D[\hat a] \hat \rho + \gamma \D[\hat a^2] \hat \rho \equiv \L \hat \rho ,\label{seqn:1}
\end{align}
with $\hat n \equiv \hat a^\dagger \hat a$. Firstly, we can observe that we have a $U(1)$ weak symmetry generated by total photon number $\hat n$. Hence, the steady state $\rho_{ss}$ defined by $\L \rho_{ss} = 0$ commutes with $\hat n$. Because the Hamiltonian $\H = \Delta \hat n + \frac{U}{2} \hat n^2$ is only a function of $\hat n$, this means the steady state is completely independent of the parameters $\Delta, U$, and so without loss of generality we can set them to be zero. Thus, we only need consider the dissipative part of the dynamics. There are two dimensionless parameters $\gamma/\kappa_g$ and $\kappa_l/\kappa_g$, and we can observe that, after optimizing over these, the fidelity to a single photon fock state never reaches above $\sim 0.6$, see \cref{fig:S3}.

\subsection{Robustness to Additional Loss}
In this section, we will consider the robustness of the topological edge mode Fock state stabilization to unwanted loss on the $B$ sublattice. We will model this as 
\begin{align}
    \dot{\hat{\rho}} &= -i[\H_{\textrm{SSH}} + \H_{\textrm{int}}, \hat \rho] + \kappa \sum_{i \in A} \D[\hat a_i] \hat \rho + \kappa \eta^2 \sum_{i \in B}  \D[\hat a_i^\dagger] \hat \rho + \gamma \sum_{i \in B} \D[\hat a_i] \hat \rho, \label{seqn:added_loss}
\end{align}
where $\H_{\textrm{SSH}}$ is defined in \cref{seqn:SSH_HAM} and $\H_{\textrm{int}} = \frac{U}{2} \sum_{i} \hat a_i^\dagger \hat a_i^\dagger \hat a_i \hat a_i$ is the same onsite Kerr nonlinearities we have been considering throughout. If we hold the nonlinearity $U = J = 1$ fixed and optimize over the remaining parameters ($\kappa, \eta, \xi$), we can see that the Fock state infidelity grows monotonically with the $B$ lattice loss rate $\gamma$ with a power law that appears to scale roughly as $(\gamma/U)^{1/3}$. 

\begin{figure}
    \centering
    \includegraphics[width=4in]{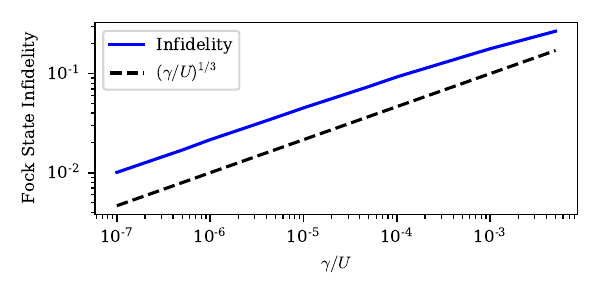}
    \caption{Fock state infidelity $1-F$ as a function of additional loss on the $B$-sublattice [c.f. \cref{seqn:added_loss}]. This shows a fixed value of the nonlinearity $U = J = 1$, and optimizing over the $A$-sublattice loss rate $\kappa$, the gain/loss ratio $\eta$, and the localization length $\xi$. The dotted line is a guide to the eye, showing power law decay $(\gamma/U)^{1/3}$.}
    \label{fig:FigS5}
\end{figure}

